\documentclass[showpacs,amsmath,amssymb,twocolumn,aps,pra,superscriptaddress,longbibliography]{revtex4-1}
\usepackage{amssymb}
\usepackage[dvips]{graphicx}
\usepackage{enumerate}
\usepackage{epsfig}
\usepackage{subfigure}
\usepackage{xcolor}
\usepackage[T1]{fontenc}
\usepackage[expert]{mathdesign}
\usepackage{fullpage}
\usepackage{amsthm,amsfonts,amssymb,amscd,mathrsfs,eufrak,xspace,framed}
\usepackage{amsmath}
\usepackage{color}
\usepackage{setspace}
\usepackage{url}
\usepackage{enumitem}
\begin{document}

\title{Twin-field Quantum Key Distribution without Phase Post-Selection}
\author{Chaohan Cui}
\author{Zhen-Qiang Yin}
\email{yinzq@ustc.edu.cn}
\author{Rong Wang}
\author{Wei Chen}
\author{Shuang Wang}
\email{wshuang@ustc.edu.cn}
\author{Guang-Can Guo}
\author{Zheng-Fu Han}
\affiliation{CAS Key Laboratory of Quantum Information, CAS Center For Excellence in Quantum Information and Quantum Physics,
University of Science and Technology of China, Hefei 230026, China}
\affiliation{State Key Laboratory of Cryptology, P. O. Box 5159, Beijing 100878, China}

\begin{abstract}
Twin-field quantum key distribution (TF-QKD) protocol and its variants, e.g. phase-matching (PM) QKD and TF-QKD based on sending or not sending, are highly attractive since they are able to overcome the well-known rate-loss limit for QKD protocols without repeater: $R=O(\eta)$ with $\eta$ standing for the channel transmittance. However, all these protocols require active phase randomization and post-selection that play an essential role together in their security proof. Counterintuitively, we find that in TF-QKD, beating the rate-loss limit is still possible even if phase randomization and post-selection in the coding mode are both removed, which means our final secure key rate $R=O(\sqrt{\eta})$. Furthermore, our protocol is more feasible in practice and more promising according to its higher final key rate in the valid distance. Our security proof counters collective attack and can also counter coherent attack in asymptotical case.

\end{abstract}

\maketitle

\section{Introduction}

With the help of quantum key distribution (QKD), two distant agents (Alice and Bob) are able to share secret key bits in the sense of information-theoretical security\cite{BB84,ShorBB84security,GLLP:2004,Rennersecurity,Scarani:QKDrev:2009,braunstein2012side,MDI}. Albeit impressive progresses on QKD experiments\cite{stucki2009high,wang20122,shibata2014quantum,pirandola2015high,korzh2015provably,yin2016measurement,yin2017satellite} have been made, there is a fundamental limit on secret key rate $R$ versus channel transmittance $\eta$. This limit is sufficiently discussed by researchers\cite{takeoka2014fundamental,pirandola2017fundamental} and finally revealed as the linear key rate bound $R\leqslant -\log_2 (1-\eta)$\cite{pirandola2017fundamental}. For a long distance, the transmittance is much smaller, then $R=O(\eta)$. Surprisingly, this limit was overcome by the twin-field (TF) QKD protocol proposed in 2018\cite{tfqkd} .
One may note that the security proof of TF-QKD has been rebuilt in Ref.\cite{Lo-tfqkd}, although its original security analysis in Ref.\cite{tfqkd} is not strict.
The physics behind TF-QKD is that Alice and Bob prepare photon-number superposition remotely via coherent states and post-selection.

Inspired by TF-QKD, phase-matching (PM) QKD protocol is introduced in Ref.\cite{pmqkd}. In PM-QKD protocol, Alice (Bob)  prepares weak coherent states $|\pm\sqrt{\mu}\rangle$ randomly and adds a random phase $\phi_A$ ($\phi_B$) to each of her (his) weak coherent states, then sends them to an untrusted party Charlie located in the middle of the channel. Depending on the measurement results declared by Charlie, Alice and Bob are able to generate raw key bits after post-selection of the cases satisfying $\phi_A\approx \phi_B$. Another variant of TF-QKD is based on sending or not sending weak coherent pulse, which can be very robust under large optical misalignment error\cite{sns-tfqkd} but the final key rate is not satisfactory. In its decoy mode, phase randomization and post-selection are still necessary. Consequently, in TF-QKD and its variants, active phase randomization and post-selection seem indispensable to the security of sifted key bits .

However, the phase post-selection may impair its secret key rate in practice. It is still an open question if the active phase randomization and phase post-selection can be removed. Here, we firstly introduce a simplified TF-QKD protocol,  in which its key bit is encoded in phase 0 or $\pi$, but unlike PM-QKD, the
coding mode does not employ active phase randomization and thus phase post-selection is also circumvented. Therefore, its coding mode is simple and the security proof is totally different from previous protocols. In section III,  the security proof of proposed protocol is given by estimating the upper bound for latent information leakage. In section IV and V, the numerical simulations with practical imperfections show that the performance of the proposed protocol without active phase randomization is satisfactory and even better, i.e., it can beat the linear key rate bound at even shorter distance than other protocols. So far we only consider threats of collective attack or coherent attack with infinite key length. A conclusion is given in section VI. 
\raggedbottom

\section{Simplified TF-QKD}Our simplified TF-QKD protocol removes the post-selection part of original TF-QKD. Firstly, let us introduce the flow of this simplified protocol as following.

Step 1. Alice and Bob randomly choose code mode or decoy mode\cite{hwang2003quantum,wang2005beating,lo2005decoy} in each trial.

Step 2.a. If code mode is selected, Alice (Bob) prepares a weak coherent state $|\pm\sqrt{\mu}\rangle_{\text{A-out}}$ ($\pm|\sqrt{\mu}\rangle_{\text{B-out}}$) according to her (his) random classical key bit $0$ or $1$, and sends the prepared state to the untrusted measurement device controlled by Eve.

Step 2.b If decoy mode is selected, Alice (Bob) emits phase-randomized weak coherent state with mean photon-number $\nu_a$ ($\nu_b$), where $\nu_a$ ($\nu_b$) is randomly chosen from a pre-decided set. Note that the phase of weak coherent state in decoy mode will be never publicly announced. Thus, in decoy mode, Alice (Bob) actually prepares a mixed state in Fock space.

Step 3. For each trial, the middle receiver Eve must publicly announce a successful message $|1\rangle_\text{M}$ or a failure message $|0\rangle_\text{M}$ to Alice and Bob. If she announces $|1\rangle_\text{M}$, she has to simultaneously declare which message she obtained, $|L\rangle_\text{M}$ or $|R\rangle_\text{M}$. For an honest Eve, $|L\rangle_\text{M}$ and $|R\rangle_\text{M}$ reveal which detector clicks\cite{pmqkd}. For simplicity, we treat the double-click event as message $|L\rangle_\text{M}$ or $|R\rangle_\text{M}$ at random. 
Note that $|1\rangle_\text{M}$, $|0\rangle_\text{M}$, $|L\rangle_\text{M}$ and $|R\rangle_\text{M}$ are all classical messages announced by Eve, though we use bra-ket notation to describe them.

Step 4. After repeating steps 1 to 3 for sufficient times, Alice and Bob publicly announce which trials are code modes and which trials are decoy modes. For the trials that Alice and Bob both select the code mode and Eve announces $|L\rangle_\text{M}$ or $|R\rangle_\text{M}$, the raw key bits are generated. Here, Bob should flip his bit if Eve announces $|R\rangle_\text{M}$. For the trials that Alice and Bob both select decoy mode, Alice and Bob can estimate the yield $Y_{n,m}$, which means the probability of Eve announcing $|1\rangle_\text{M}$ provided Alice emits $n$-photon state and Bob emits $m$-photon state in a decoy mode. With these parameters, information leakage is bounded so that secret key bits can be generated from raw key bits by error correction and privacy amplification.

In the following paper, we will focus on the upper bound for the information leakage through the whole protocol.

\section{Main results of security Proof}
For readability, we sketch the security proof and its main results here. One may refer to Appendix A for detailed derivations.
 We make no more assumptions to Eve than assumptions applied in measurement-device-independent (MDI) QKD\cite{MDI,braunstein2012side}. Accordingly, Eve's general collective attack to the above simplified TF-QKD protocol can be defined as an arbitrary measurement after an arbitrary unitary operation operating on the whole system with her prepared ancilla\cite{Rennersecurity,Scarani:QKDrev:2009}. Under photon-number representation, this collective attack is given by
\begin{equation}
\begin{aligned}
\label{attack-text}
&\hat U|n\rangle_{\text{A-out}}|m\rangle_{\text{B-out}}|E_0\rangle_\text{Ea}|0\rangle_\text{M} \\
&= \sqrt{Y_{n,m}}|\gamma_{n,m}\rangle_\text{E} |1\rangle_\text{M} + \sqrt{1-Y_{n,m}}|\text{other}\rangle_\text{E} |0\rangle_\text{M},
\end{aligned}
\end{equation}
where $|n\rangle_{\text{A-out}}$ and $|m\rangle_{\text{B-out}}$ represent the photon-number bases of the quantum states prepared by Alice and Bob respectively, 
the state $|E_0\rangle_{\text{Ea}}$ is the ancilla of Eve, and $Y_{n,m}\in[0,1]$ is a probability-like value shows the portion that Alice and Bob receive the message $|1\rangle_\text{M}$ form Eve. On the right side of Eq.\eqref{attack-text}, $|\gamma_{n,m}\rangle_\text{E}$ and $|\text{other}\rangle_\text{E}$ are the quantum states of compound system including Eve's ancilla Ea, A-out and B-out, which are all in the hands of Eve now. Note that any phases of the states on the right hand side of Eq.\eqref{attack-text} are absorbed
into the definition of those states. For simplicity, let's denote Eve's message $|L\rangle_\text{M}$ and $|R\rangle_\text{M}$ as the same one $|1\rangle_\text{M}$, since we only concern Alice's key bit here, but not Bob's bit and his flipping operation. We aim to bound Eve's information $I_\text{AE}$ on Alice's key bit when Eve announces message $|1\rangle_\text{M}$. Through derivations given in the Appendix A, it is proved that this upper bound $I^{u}_\text{AE}$ can be solved by the following optimization problem given by
\begin{equation}
\label{Iae}
\begin{aligned}
&I^{u}_\text{AE}
=\text{max}\,\,  h(\dfrac{x_{00}}{Q_\mu},\dfrac{x_{10}}{Q_\mu})+h(\dfrac{x_{11}}{Q_\mu},\dfrac{x_{01}}{Q_\mu}),\\
&s.t.\quad
\begin{cases}
0\leqslant x_{00}\leqslant\big|\sum_{n,m=0}\sqrt{P_{2n}P_{2m}Y_{2n,2m}} \big|^2, \\
0\leqslant x_{10}\leqslant\big|\sum_{n,m=0}\sqrt{P_{2n+1}P_{2m}Y_{2n+1,2m}} \big|^2,\\
0\leqslant x_{11}\leqslant\big|\sum_{n,m=0}\sqrt{P_{2n+1}P_{2m+1}Y_{2n+1,2m+1}} \big|^2,\\
0\leqslant x_{01}\leqslant\big|\sum_{n,m=0}\sqrt{P_{2n}P_{2m+1}Y_{2n,2m+1}} \big|^2,\\
x_{00}+x_{10}+x_{11}+x_{01}=Q_{\mu}.
\end{cases}
\end{aligned}
\end{equation}
with the definition $h(x,y)=-x\log_2x-y\log_2y+(x+y)\log_2 (x+y)$. Here, $P_{k}=e^{-\mu}\mu^k/k!$ is the probability of coherent state $\pm|\sqrt{\mu}\rangle$ containing $k$-photons, and $Q_\mu$ is the probability of Alice obtaining a raw key bit in code mode, which is directly observed experimentally.
In practice, agents can observe the parameters $P_n$, $P_m$, $Q_{\mu}$, and $Y_{n,m}$. Then, the information leakage bound can be estimated by the above optimization problem. According to Devetak-Winter's bound \cite{devetak2005distillation}, the secret key rate per trial in a code mode is then given by
\begin{equation}\begin{aligned}
\label{KeyRate-text}
R = Q_\mu \big[1-f h(e_\mu,1-e_\mu)-I^{u}_\text{AE}],
\end{aligned}\end{equation}
in which, $e_\mu$ is the error rate of raw key bits. This security proof assumes that Eve only launches collective attack, however, this restriction can be removed by following the results in Refs \cite{caves2002unknown,christandl2009postselection}. Hence, our proof can guarantee the security against the coherent attacks asymptotically. It also ends our security proof rigorously.

Before proceeding, let's roughly estimate the performance of the protocol under ideal case, in which only channel transmission efficiency $\eta$ is considered, while all other imperfections, e.g. dark counts of single photon detectors, are absent. Then, it is expected that $x_{00} \sim  x_{11} \sim \mu^2 O(\sqrt{\eta})$, since the main contribution of $x_{00}$ and $x_{11}$ comes from the yield of the total photon number from Alice and Bob is two. With similar argument, we have $x_{01} \sim  x_{10} \sim \mu O(\sqrt{\eta})$. Thus, from Eq.\eqref{Iae} we can see $I^{u}_\text{AE}\ll 1$ for any $\eta$ provided a proper value of $\mu$ is assumed.
Besides, it is obvious that $Q_\mu \sim \mu O(\sqrt{\eta})$ and $e_\mu = 0$ in ideal case. Accordingly, from above formulae, we have $R = O (\sqrt{\eta})$. This does reconfirm the expectation that the TF-QKD can overcome linear bound even if phase randomization and post-selection are both removed.
In the next two sections, through numerical simulations with practical imperfections we will show the performance of our protocol with both infinite and finite decoy states techniques comparing with other states of the art.

\section{Estimation and Simulation with Infinite Decoy States}

In a practical system, Alice and Bob can emit phase randomized decoy states\cite{hwang2003quantum} to estimate $Y_{n,m}$. The gain of the decoy states that Alice emits pulse with mean photon-number $\nu_a$ and Bob emits pulse with mean photon-number $\nu_b$ shall satisfy
\begin{equation}
\label{infdcy}
\begin{aligned}
&Q^{\nu_a,\nu_b}_\text{d}=\sum_{n,m}P_{n}^{\nu_a}P_{m}^{\nu_b} Y_{n,m},
\end{aligned}\end{equation}
where $P_{k}^{\nu}=e^{-\nu}\nu^k/k!$ is known by both agents.
Considering the ideal case with infinite decoy states $\nu_a$ and $\nu_b$, we can list infinite linear equations like Eq.\eqref{infdcy} to calculate $Y_{n,m}$ accurately. Therefore, the secure key rate can be easily calculated by Eq.\eqref{KeyRate-text} with $I^{u}_\text{AE}$ given by Eq.\eqref{Iae}. Here we simulate the maximum secure key rates related to different loss for multiple protocol with infinite decoy states implement and practical parameter of experiments. Details can be found in the Appendix B. The results are shown in Fig.\ref{Fig1}.
\begin{figure*}[htbp]
\begin{tabular}{ll}
\begin{minipage}[h]{0.6\textwidth}
\centering
\includegraphics[width=10cm]{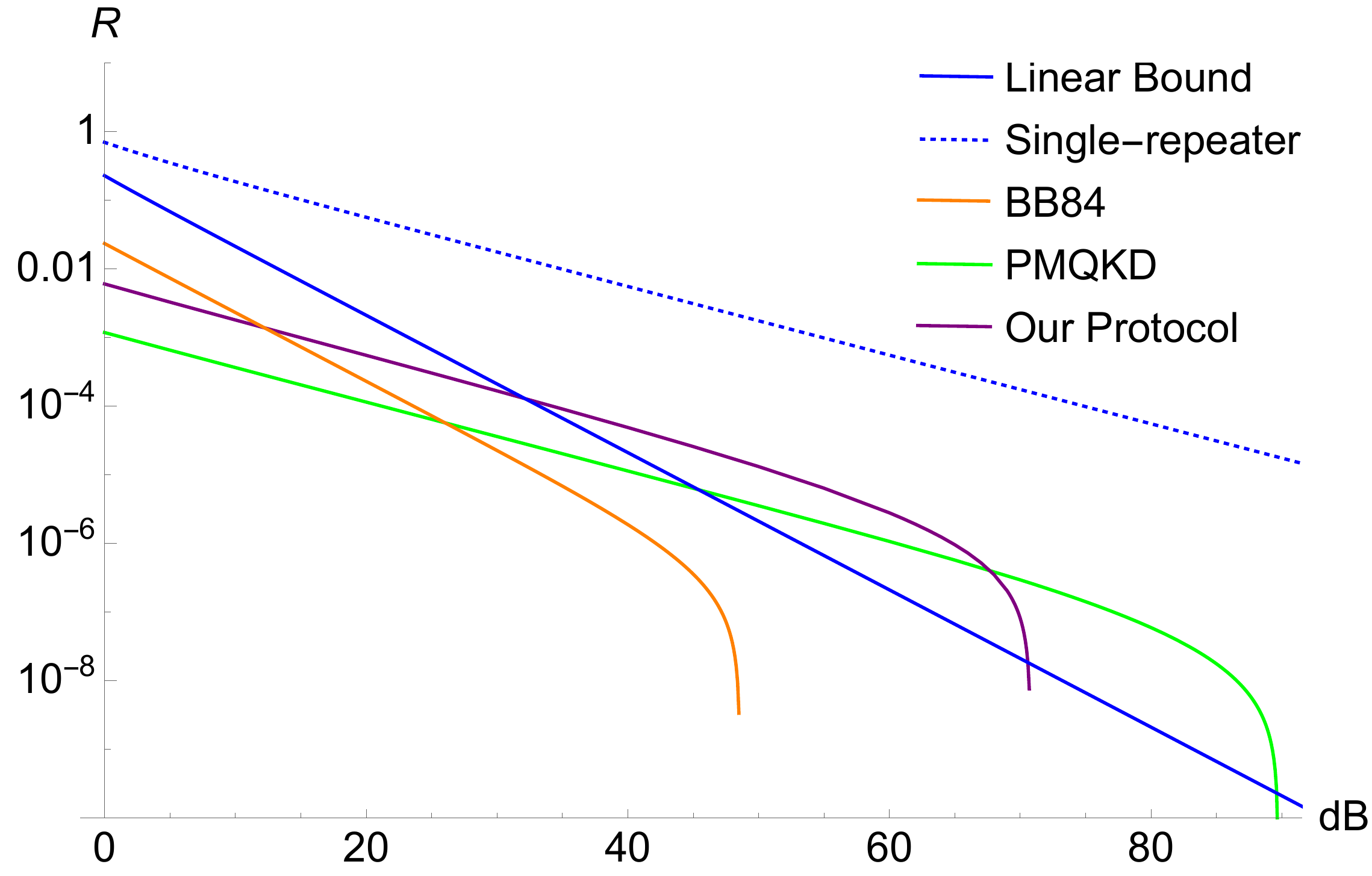}
\end{minipage} &
\begin{minipage}[h]{0.4\textwidth}
\centering
\begin{tabular}{ll}
\hline
Parameters & Values \\
\hline
Dark count rate $p_d$ & $8\times10^{-8}$\\
Error correction efficiency $f$ & 1.15 \\
Detector efficiency $\eta_d$ & 14.5\%\\
Misalignment error $e_M$ &  0.375\%\\
\hline
\end{tabular}
\end{minipage}
\end{tabular}
\caption{Key rates(R) of our protocol(purple), PM-QKD(green)\cite{pmqkd} and BB84(orange) with all infinite decoy states versus different optical fiber losses (dB). Note that we assume zero optical misalignment, except the misalignment due to phase post-selection ($M=16$) of original PM-QKD in Ref.\cite{pmqkd}. Linear key rate bound\cite{pirandola2017fundamental}(blue) and it with single repeater\cite{pirandola2016capacities}(blue, dotted) are also shown in the figure.}
\label{Fig1}
\end{figure*}
We can see that with infinite decoy states, our protocol has higher key rate than original PM-QKD because our protocol is phase post-selection free and independent with extra error estimations. Note that the slope of key rate in our protocol is the same as linear bound with single repeater \cite{pirandola2016capacities} when the fiber loss is less than 60 dB, which shows that advantage of beating well-known linear bound is also reconfirmed through 30 dB to 60 dB fiber loss. In other words, $R = O(\sqrt{\eta})$. It is also remarkable that our protocol can outperform BB84 at lower channel loss comparing to the original PM-QKD.

\section{Estimation and Simulation with Finite Decoy States}
Finite decoy states can also help to estimate the lower bound for yields\cite{wang2005beating,lo2005decoy,zhou2016making}. In a practical system, this implement is much more feasible than the infinite one. Here we apply decoy states with four different intensities as $\mu$, $\nu_1$, $\nu_2$ and $0$. After announcement of decoy modes and each applied intensity, we have gains as $Q_d^{0,0}$, $Q_d^{\mu,0}$, $Q_d^{\nu_1,0}$, $Q_d^{\nu_2,0}$, $Q_d^{0,\mu}$, $Q_d^{0,\nu_1}$, $Q_d^{0,\nu_2}$, $Q_d^{\mu,\mu}$, $Q_d^{\nu_1,\nu_1}$ and $Q_d^{\nu_2,\nu_2}$.

Then we show how those statistics can give good approximations to $Y_{0,0}$, $Y_{0,1}$, $Y_{1,0}$, $Y_{2,0}$, $Y_{0,2}$ and $Y_2$, where $Y_2$ means the yield that are from decoy trials in which Alice and Bob share $2$ photons in total. From $Q_d^{\mu,0}$, $Q_d^{\nu_1,0}$, $Q_d^{\nu_2,0}$ and $Q_d^{0,0}$, we can obtain lower bounds and upper bounds of $Y_{0,0}$, $Y_{0,1}$, $Y_{0,2}$ by linear programming on Eq.\eqref{infdcy}. Similarly, lower bounds and upper bounds of $Y_{1,0}$ and $Y_{2,0}$ can be estimated from $Q_d^{0,\mu}$, $Q_d^{0,\nu_1}$, $Q_d^{0,\nu_2}$ and $Q_d^{0,0}$. Upper bound and lower bound of $Y_2$ could also be bounded by the linear programming on four linear equations of $Q_d^{\mu,\mu}$, $Q_d^{\nu_1,\nu_1}$, $Q_d^{\nu_2,\nu_2}$ and $Q_d^{0,0}$. In the following text, we use superscript $u$ or $l$ to label the upper or lower bound for $Y$ obtained here. To estimate $Y_{1,1}$, through the relation $Y_2=\dfrac{\sum_i P^\mu_{i}P^\mu_{2-i}Y_{i,2-1}}{\sum_i P^\mu_{i}P^\mu_{2-i}}$, $Y_{1,1}$ could be bounded by
\begin{widetext}
\begin{equation}\begin{aligned}
\label{conY11}
\dfrac{(2P^\mu_{0}P^\mu_{2}+P^\mu_{1}P^\mu_{1})Y_2^l-P^\mu_{0}P^\mu_{2}(Y^u_{0,2}+Y^u_{2,0})}{P^\mu_{1}P^\mu_{1}} = Y_{1,1}^l \leqslant Y_{1,1}\leqslant Y_{1,1}^u = \dfrac{(2P^\mu_{0}P^\mu_{2}+P^\mu_{1}P^\mu_{1})Y_2^u-P^\mu_{0}P^\mu_{2}(Y^l_{0,2}+Y^l_{2,0})}{P^\mu_{1}P^\mu_{1}}
\end{aligned}\end{equation}\end{widetext}
Then the remained task is constraining $x_{00}$, $x_{01}$, $x_{10}$ and $x_{11}$ with these lower bounds and upper bounds generated from decoy statistics.

Let's take $x_{00}$ as an example. From Eq.\eqref{Iae}, we have a general bound that limits the $x_{00}$ as
\begin{widetext}\begin{equation}\begin{aligned}
\label{psiee}
&x_{00}\leqslant \big|\sum_{n,m=0}\sqrt{P^\mu_{2n}P^\mu_{2m}Y_{2n,2m}} \big|^2\leqslant \max_{k\geqslant 2}\big|\sqrt{P^\mu_{0}P^\mu_{0}Y^u_{0,0}}+\sqrt{P^\mu_{0}P^\mu_{2}Y^u_{0,2}}+\sqrt{P^\mu_{2}P^\mu_{0}Y^u_{2,0}}+\\
&\sqrt{\dfrac{(k+4)(k-1)}{2}(Q_d^{\mu,\mu}-\sum_{n=k+1}^{+\infty}\sum_{i=0}^nP^\mu_{2i}P^\mu_{2(n-i)}-\sum_{n=0}^{2}\sum_{i=0}^nP^\mu_{i}P^\mu_{n-i}Y^l_{i,n-i})} + \sum_{n=k+1}^{+\infty}\sum_{i=0}^n\sqrt{P^\mu_{2i}P^\mu_{2(n-i)}}\big|^2
\end{aligned}\end{equation}\end{widetext}
The details of the derivation are included in the appendix B. Now we obtain the bounds of these four values based on all the observables in our protocol. The final step is only making an optimization to find the best information-theoretical secure key rate with Eq.\eqref{KeyRate-text} limited by these bounds.

\begin{figure*}[htbp]
\begin{tabular}{ll}
\begin{minipage}[h]{0.6\textwidth}
\centering
\includegraphics[width=10cm]{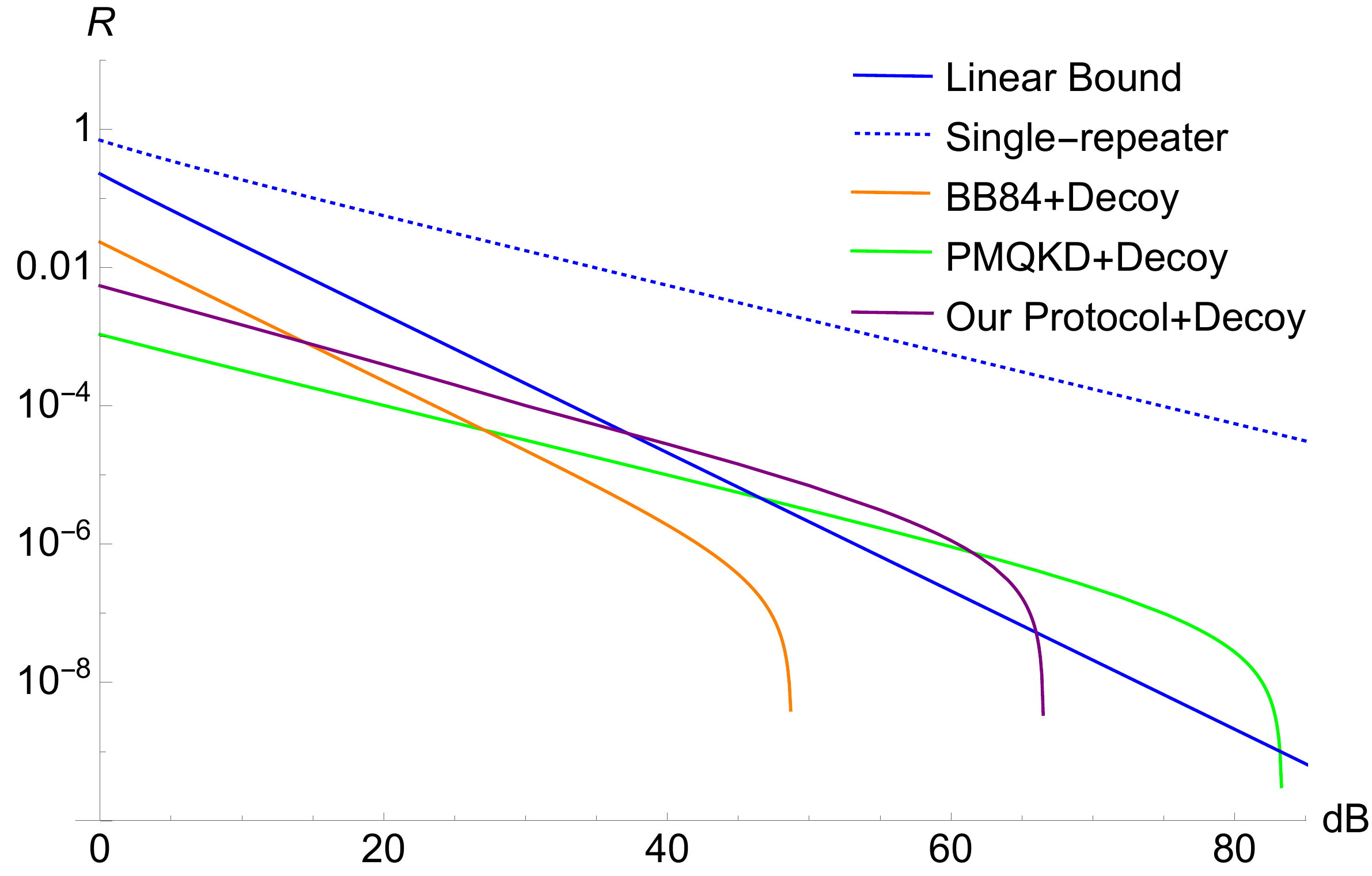}
\end{minipage} &
\begin{minipage}[h]{0.4\textwidth}
\centering
\begin{tabular}{ll}
\hline
Parameters & Values \\
\hline
Dark count rate $p_d$ & $8\times10^{-8}$\\
Error correction efficiency $f$ & 1.15 \\
Detector efficiency $\eta_d$ & 14.5\%\\
Misalignment error $e_M$ &  0.375\%\\
\hline
\end{tabular}
\end{minipage}
\end{tabular}
\caption{Key rates(R) of our protocol(purple), PM-QKD(green)\cite{pmqkd} and BB84(orange) with all finite decoy states versus different optical fiber losses (dB). Note that we assume zero optical misalignment, except the misalignment due to phase post-selection ($M=16$) of original PM-QKD in Ref.\cite{pmqkd}. Linear key rate bound\cite{pirandola2017fundamental}(blue) and it with single repeater\cite{pirandola2016capacities}(blue, dotted) are also shown in the figure.}
\label{Fig2}
\end{figure*}

So far, we show how Alice and Bob can estimate the lower bound for the key rates under different losses with four decoy states . We simulate a practical case for multiple protocols. The results are shown in Fig.\ref{Fig2}. Even in the case of finite decoy states, our protocol's key rate holds the relation with transmittance as $R=O(\sqrt{\eta})$. Consequently, it can still beat the linear bound in the loss range from 40 dB to 60 dB.

\section{Conclusion}
Inspired by TF-QKD protocol and its variants such as PM-QKD, we proposed a simplified protocol with higher final key rate, in which the raw key bits are generated without active phase randomization and phase post-selection. A meticulous security proof is presented by estimating the information leakage in our protocol. Counterintuitively, our bound for latent information leakage doesn't rely on the error rate. Meanwhile, its advantage of beating the linear rate-loss limit is still available here, showing that the final key rate $R=O(\sqrt{\eta})$ over transmittance $\eta$. Besides, thanks to the removal of phase post-selection, our scheme can perform over the well-known BB84 at a shorter channel distance comparing to original PM-QKD protocol, which means the proposed protocol could be very competitive when channel loss is around $15$dB to $60$dB. 

\emph{Note added.} After posting our work on arXiv, two other groups provided similar ideas \cite{curty2018simple,lin2018simple} independently but didn't include the practical case with finite decoy states. In methodology, our work is based on the analysis of collective attack while the Ref.\cite{curty2018simple} presents a proof based on an equivalent entanglement distillation protocol. 
Besides, both the Refs.\cite{curty2018simple,lin2018simple} are using infinite decoy states which is not feasible in the experiments. 

\section{Acknowledgement}
This work has been supported by the National Key Research and Development Program of China (Grant No. 2016YFA0302600), the National Natural Science Foundation of China (Grant Nos. 61822115, 61775207, 61622506, 61627820, 61575183), Anhui Initiative in Quantum Information Technologies.

\bibliography{MainBib}

\onecolumngrid
\appendix*
\section*{Appendix A: Security proof}
We make no more assumptions to Eve than assumptions applied in measurement-device-independent (MDI) QKD\cite{MDI,braunstein2012side}. In order to bounding the information leakage to Eve, we have to describe the ultimate power of Eve under the assumptions. Also, Eve's strategy must obey the time line through this protocol. Therefore, Eve's general collective attack to the above simplified TF-QKD protocol can be defined as an arbitrary measurement after an arbitrary unitary operation operating on the whole system with her prepared ancilla\cite{Rennersecurity,Scarani:QKDrev:2009}. Furthermore, the message Eve announces should be also obtained from the measurement results. Under photon-number representation, this collective attack is given by

\begin{equation}
\label{attack}
\hat U|n\rangle_{\text{A-out}}|m\rangle_{\text{B-out}}|E_0\rangle_\text{Ea}|0\rangle_\text{M} = \sqrt{Y_{n,m}}|\gamma_{n,m}\rangle_\text{E} |1\rangle_\text{M} + \sqrt{1-Y_{n,m}}|\text{other}\rangle_\text{E} |0\rangle_\text{M},
\end{equation}
where $|n\rangle_{\text{A-out}}$ and $|m\rangle_{\text{B-out}}$ represent the photon-number bases of the quantum states prepared by Alice and Bob respectively, 
the state $|E_0\rangle_{\text{Ea}}$ is the ancilla of Eve, and $Y_{n,m}\in[0,1]$ is a probability-like value shows the portion that Alice and Bob receive the message $|1\rangle_\text{M}$ form Eve. On the right side of Eq.\eqref{attack}, $|\gamma_{n,m}\rangle_\text{E}$ and $|\text{other}\rangle_\text{E}$ are the quantum states of compound system including Eve's ancilla Ea, A-out and B-out, which are all in the hands of Eve. Note that any phases of the states on the right hand side of Eq.\eqref{attack} are absorbed
into the definition of those states. For simplicity, let's denote Eve's message $|L\rangle_\text{M}$ and $|R\rangle_\text{M}$ as the same one $|1\rangle_\text{M}$, since we only concern Alice's key bit here, but not Bob's bit and his flipping operation. Note that this expression does give the most general collective attack, including possible attacks trying to distinguish decoy mode and code mode and treat them differently, since Eve's ancilla is arbitrary. Indeed, any measurement and following transformation depending on the output of the measurement can be described as a "giant" unitary operator applied to a larger Hilbert space.

Suppose Alice and Bob each has an ancillary qubit to store their classical key bit in code mode. For simplicity, here we assume that Alice and Bob's random binary bits come from measurements of their qubits in $Z$ bases. So they set their initial qubits to $|+\rangle$ and prepare a weak coherent state light pulse with average photon-number $\mu$.  Then the initial prepared state is
\begin{equation}
|+\rangle_{\text{A}}|+\rangle_{\text{B}}|\sqrt{\mu}\rangle_{\text{A-out}}|\sqrt{\mu}\rangle_{\text{B-out}}.
\end{equation}

Then Alice and Bob apply a C-$\pi$ gate to upload their information on the output coherent state and measure their private qubits. Recall Eve's attack given by Eq.\eqref{attack}. For the ease of representation, we define four intermediate unnormalized states labeled by the photon-number's parity of A-out and B-out,
\begin{equation}
\begin{aligned}
\label{psi}
&|\psi_{ee}\rangle=\sum_{n,m}\sqrt{P_{2n}P_{2m}Y_{2n,2m}}|\gamma_{2n,2m}\rangle,\ \ |\psi_{oo}\rangle=\sum_{n,m}\sqrt{P_{2n+1}P_{2m+1}Y_{2n+1,2m+1}}|\gamma_{2n+1,2m+1}\rangle,\\
&|\psi_{eo}\rangle=\sum_{n,m}\sqrt{P_{2n}P_{2m+1}Y_{2n,2m+1}}|\gamma_{2n,2m+1}\rangle,\ \ |\psi_{oe}\rangle=\sum_{n,m}\sqrt{P_{2n+1}P_{2m}Y_{2n+1,2m}}|\gamma_{2n+1,2m}\rangle,
\end{aligned}
\end{equation}
where the subscript E is dropped for simplicity.
Since Alice and Bob's encoding phases are $0$ or $\pi$, the phase of Fock state $|n\rangle_{\text{A-out}}|m\rangle_{\text{B-out}}$ will not change if both $n$ and $m$ are odd or even, while $|n\rangle_{\text{A-out}}|m\rangle_{\text{B-out}}$ will change to $-|n\rangle_{\text{A-out}}|m\rangle_{\text{B-out}}$ if only one of $n$ and $m$ is odd. Thus in code mode, under the same combination of coding phases, the phase of superposition of $|n\rangle_{\text{A-out}}|m\rangle_{\text{B-out}}$ can be divided into four groups depending the parity of $n$ and $m$. This implies we can define Eq.\eqref{psi}, which is just the superposition of $|n\rangle_{\text{A-out}}|m\rangle_{\text{B-out}}$ with different parities. Those states $|\psi_{ee}\rangle$, $|\psi_{oo}\rangle$, $|\psi_{eo}\rangle$ and $|\psi_{oe}\rangle$ are quite useful for simplifying the following derivations. 

It should be taken into account that Eq.\eqref{attack} never implies whether $|\gamma_{n,m}\rangle$ are orthogonal to each other or not. It's obvious that Alice and Bob can not obtain any direct knowledge of them because they are measured by Eve. After tracing Bob's qubit out and measuring Alice's qubit in $Z$ basis, the unnormalized density matrix of Eve's system E and mode A conditioned that $|1\rangle_M$ is announced becomes
\begin{equation}\begin{aligned}
\label{rho_ae}
\rho_\text{AE}
=&\dfrac{1}{2}\Pi\{|0\rangle_{\text{A}}\}\otimes(\Pi\{|\psi_{ee}\rangle+|\psi_{oe}\rangle\}+\Pi\{|\psi_{oo}\rangle+|\psi_{eo}\rangle\})\\
+&\dfrac{1}{2}\Pi\{|1\rangle_{\text{A}}\}\otimes(\Pi\{|\psi_{ee}\rangle-|\psi_{oe}\rangle\}+\Pi\{|\psi_{oo}\rangle-|\psi_{eo}\rangle\}),
\end{aligned}\end{equation}
where, $\Pi\{|\psi\rangle\}=|\psi\rangle\langle\psi|$. Then the Holevo bound of $\rho_\text{AE}$ is upper-bounded by
\begin{equation}
\begin{aligned}
\label{chi}
\chi(\rho_\text{AE})\leqslant &\dfrac{h(\big||\psi_{ee}\rangle\big|^2,\big||\psi_{oe}\rangle\big|^2)+h(\big||\psi_{oo}\rangle\big|^2,\big||\psi_{eo}\rangle\big|^2)}{\big||\psi_{ee}\rangle\big|^2+\big||\psi_{oe}\rangle\big|^2+\big||\psi_{oo}\rangle\big|^2+\big||\psi_{eo}\rangle\big|^2}=h(\dfrac{\big||\psi_{ee}\rangle\big|^2}{Q_\mu},\dfrac{\big||\psi_{oe}\rangle\big|^2}{Q_\mu})+h(\dfrac{\big||\psi_{oo}\rangle\big|^2}{Q_\mu},\dfrac{\big||\psi_{eo}\rangle\big|^2}{Q_\mu})
\end{aligned}
\end{equation}
with the definition $h(x,y)=-x\log_2x-y\log_2y+(x+y)\log_2 (x+y)$, which is different from the definition of binary Von Neumann entropy. To get the inequality of Eq.\eqref{chi}, we first note that Eve's system E is a mixture of $\Pi\{|\psi_{ee}\rangle+|\psi_{oe}\rangle\}$, $\Pi\{|\psi_{ee}\rangle-|\psi_{oe}\rangle\}$, $\Pi\{|\psi_{oo}\rangle+|\psi_{eo}\rangle\}$ and $\Pi\{|\psi_{oo}\rangle-|\psi_{eo}\rangle\}$. Moreover, without compromising the security, one may assume Eve gets some side-channel information or a partial purification of $\rho_\text{E}$, which just honestly tells Eve that her system is one of $\Pi\{|\psi_{ee}\rangle+|\psi_{oe}\rangle\}$ and $\Pi\{|\psi_{ee}\rangle-|\psi_{oe}\rangle\}$, or one of $\Pi\{|\psi_{oo}\rangle+|\psi_{eo}\rangle\}$ and $\Pi\{|\psi_{oo}\rangle-|\psi_{eo}\rangle\}$. This assumption just helps Eve to guess Alice's key bit and simplifies the calculation of $\chi(\rho_\text{AE})$ greatly. 
The probability of Alice obtaining a raw key bit in a code mode can be presented by $Q_\mu=\big||\psi_{ee}\rangle\big|^2+\big||\psi_{oe}\rangle\big|^2+\big||\psi_{eo}\rangle\big|^2+\big||\psi_{oo}\rangle\big|^2$.  Now, we have clearly show that Eve's information on Alice's classical key bit is bounded by Eq.\eqref{chi} as $I(A:E)\leqslant\chi(\rho_\text{AE})$, even when the active phase randomization is removed. According to Devetak-Winter's bound \cite{devetak2005distillation},  the secret key rate per trial in code mode is
\begin{equation}\begin{aligned}
\label{KeyRate}
R = Q_\mu \big[1-f h(e_\mu,1-e_\mu)-h(\dfrac{\big||\psi_{ee}\rangle\big|^2}{Q_\mu},\dfrac{\big||\psi_{oe}\rangle\big|^2}{Q_\mu})-h(\dfrac{\big||\psi_{oo}\rangle\big|^2}{Q_\mu},\dfrac{\big||\psi_{eo}\rangle\big|^2}{Q_\mu})],
\end{aligned}\end{equation}
in which, $e_\mu$ is the error rate of raw key bits. To calculate $\chi(\rho_\text{AE})$, these four values $\big||\psi_{ee}\rangle\big|^2$, $\big||\psi_{oo}\rangle\big|^2$, $\big||\psi_{oe}\rangle\big|^2$ and $\big||\psi_{eo}\rangle\big|^2$ must be estimated. Obviously, Alice and Bob can relate these values to the direct observables and statistics, i.e. 
\begin{equation}
\begin{aligned}
\label{cons}
&\big||\psi_{ee}\rangle\big|^2\leqslant\big|\sum_{n,m=0}\sqrt{P_{2n}P_{2m}Y_{2n,2m}} \big|^2, \\
&\big||\psi_{oe}\rangle\big|^2\leqslant\big|\sum_{n,m=0}\sqrt{P_{2n+1}P_{2m}Y_{2n+1,2m}} \big|^2,\\
&\big||\psi_{oo}\rangle\big|^2\leqslant\big|\sum_{n,m=0}\sqrt{P_{2n+1}P_{2m+1}Y_{2n+1,2m+1}} \big|^2,\\
&\big||\psi_{eo}\rangle\big|^2\leqslant\big|\sum_{n,m=0}\sqrt{P_{2n}P_{2m+1}Y_{2n,2m+1}} \big|^2,\\
&\big||\psi_{ee}\rangle\big|^2+\big||\psi_{oe}\rangle\big|^2+\big||\psi_{oo}\rangle\big|^2+\big||\psi_{eo}\rangle\big|^2=Q_{\mu}.
\end{aligned}
\end{equation}
With these constraints, one can estimate upper bound of $\chi(\rho_\text{AE})$. By defining $x_{00} \triangleq  \big||\psi_{ee}\rangle\big|^2$, $x_{10} \triangleq  \big||\psi_{oe}\rangle\big|^2$, $x_{11} \triangleq  \big||\psi_{oo}\rangle\big|^2$, and $x_{01} \triangleq  \big||\psi_{eo}\rangle\big|^2$, we reach the Eq.\eqref{Iae} in the main text.

\section*{Appendix B: Details of mathematics in simulation}

We derive a simulation scheme for our protocol and give out numerical results. Suppose the dark count of each detector is $p_d$ per trial and each partner sends a coherent state carrying average $\mu$ photons. After the lossy channel and the interference of Alice and Bob's pulse, the coherent state flows to the correct detector with zero misalignment of devices and no attack. However, due to the loss and the dark count, the response probability is less than 1 and may come from the wrong detector. For each trial, the correct case is when the correct detector provides a response (no matter whether it comes from a dark count or a real signal) and simultaneously there's no dark count from another detector. The probability $P_{c}=(1-p_d)[1-(1-p_d)\exp(-2\eta\mu)]$. Also, an error case occurs when the wrong detector clicks for a dark count while the correct detector gets nothing, with a probability $P_{e}=(1-p_d)\exp(-2\eta\mu)p_d$. The gain $Q_\mu$ of code mode should be
\begin{equation}
Q_\mu=P_{c}+P_{e}=(1-p_d)[1-(1-p_d)\exp(-2\eta\mu)]+(1-p_d)\exp(-2\eta\mu)p_d
\end{equation}
The error rate should be
\begin{equation}
e_\mu=\dfrac{P_{e}}{P_{c}+P_{e}}=\dfrac{\exp(-2\eta\mu)p_d}{1-(1-2p_d)\exp(-2\eta\mu)}
\end{equation}
The above formulas are in accord with results of Ref.\cite{pmqkd} with zero misalignment. The misalignment can be also included both in the gain and error rate, but in our phase-randomization-free protocol we assume the best performance that the misalignment is zero.

If we apply infinite decoy states, the approximation of $Y_{n,m}$ can be calculate as
\begin{equation}
Y_{n,m} = 1-(1-p_d)^2(1-\eta)^{n + m}.
\end{equation}
Here, without compromising security, we treat double-click event as message $|L\rangle_\text{M}$ or $|R\rangle_\text{M}$ at random to simplify the bound of $Y_{n,m}$. With the above equations, Eq.\eqref{cons} is bounded by  parameters in real experiments.

If we apply finite decoy states, we can only obtain good bounds for several $Y_{n,m}$ with small $n$ and $m$. For the case considered in the main text, linear programing on statistics and Eq.\eqref{conY11} helps to bound $Y_{0,0}$, $Y_{0,1}$, $Y_{1,0}$, $Y_{2,0}$, $Y_{0,2}$ and $Y_{1,1}$.
The upper bound for the right-hand values in Eq.\eqref{cons} could be calculated as an optimization problem with constrains. Recall the example in the main text.
\begin{equation}\begin{aligned}
\label{psiee2}
x_{00}&\leqslant \big|\sum_{n,m=0}^{+\infty}\sqrt{P^\mu_{2n}P^\mu_{2m}Y_{2n,2m}} \big|^2=\big|\sum_{n=0}^{+\infty}\sum_{i=0}^n\sqrt{P^\mu_{2i}P^\mu_{2(n-i)}Y_{2i,2(n-i)}} \big|^2\\
&\leqslant\big|\sqrt{P^\mu_{0}P^\mu_{0}Y^u_{0,0}}+\sqrt{P^\mu_{0}P^\mu_{2}Y^u_{0,2}}+\sqrt{P^\mu_{2}P^\mu_{0}Y^u_{2,0}}+\sum_{n=2}^{+\infty}\sum_{i=0}^n\sqrt{P^\mu_{2i}P^\mu_{2(n-i)}Y_{2i,2(n-i)}} \big|^2\\
\end{aligned}
\end{equation}
$\sum_{n=2}\sum_{i=0}^n\sqrt{P^\mu_{2i}P^\mu_{2(n-i)}Y_{2i,2(n-i)}}$ is untouchable by statistics from only four decoy states. But the constrains of total gain give an upper bound for this term. The optimization problem can be described as
\begin{equation}
\label{cons2}
\begin{aligned}
&\max\,\, \sum_{n=2}^{+\infty}\sum_{i=0}^n\sqrt{P^\mu_{2i}P^\mu_{2(n-i)}Y_{2i,2(n-i)}},\\
&s.t.\quad
\begin{cases}
\sum\limits_{n=2}^{+\infty}\sum\limits_{i=0}^nP^\mu_{2i}P^\mu_{2(n-i)}Y_{2i,2(n-i)} \leqslant Q_d^{\mu,\mu}- \sum\limits_{n=0}^2\sum\limits_{i=0}^nP^\mu_{i}P^\mu_{n-i}Y^l_{i,n-i}, \\
0\leqslant Y_{i,j} \leqslant 1 \,\, \forall  i,j\in \mathbb{Z},
\end{cases}
\end{aligned}
\end{equation}
which could be easily solved numerically. On the other side, we can use an analytical approach for the upper bound of the aimed function in Eq.\eqref{cons2}. Since all known probability terms are positive and decrease to zero exponentially, the above optimization problem satisfies the well-known Karush-Kuhn-Tucker \cite{boyd2004convex} conditions. Therefore, the maximum is located at the boundary where any $Y_{q,t-q}$ with $t>k$ reaches its upper bound and others hold the conditions. Then the problem is simplified to finding an integer $k\geqslant2$ that reaches the maximum value of the aimed function in Eq.\eqref{cons2}, say
\begin{equation}
\label{cons3}
\sum_{n=2}^\infty\sum_{i=0}^n\sqrt{P^\mu_{2i}P^\mu_{2(n-i)}Y_{2i,2(n-i)}} \leqslant \max_{k\geqslant 2}\,\, \sum_{n=2}^k\sqrt{(n+1)\sum_{i=0}^nP^\mu_{2i}P^\mu_{2(n-i)}Y_{2i,2(n-i)}}+\sum_{n=k+1}^\infty\sum_{i=0}^n\sqrt{P^\mu_{2i}P^\mu_{2(n-i)}}.
\end{equation}
Here at the right-hand side, $k\geqslant 2$ is an integer waiting for optimization. This inequality is based on the inequality between arithmetic mean and quadratic mean. Then, the optimization problem Eq.\eqref{cons2} becomes
\begin{equation}
\begin{aligned}
&\max_{k\geqslant 2}\,\, \sum_{n=2}^k\sqrt{(n+1)\alpha_n}+\sum_{n=k+1}^\infty\sum_{i=0}^n\sqrt{P^\mu_{2i}P^\mu_{2(n-i)}},\\
&s.t.\quad
\begin{cases}
\sum\limits_{n=2}^k \alpha_n \leqslant Q_d^{\mu,\mu}- \sum\limits_{n=0}^2\sum\limits_{i=0}^nP^\mu_{i}P^\mu_{n-i}Y^l_{i,n-i}, \\
0\leqslant \alpha_n \leqslant\sum\limits_{i=0}^nP^\mu_{2i}P^\mu_{2(n-i)} \, , \, \forall \,\,  2\leqslant n\leqslant k.
\end{cases}
\end{aligned}
\end{equation}

\begin{equation}
\begin{aligned}
\label{cons4}
&\sum_{n=2}^{+\infty}\sum_{i=0}^n\sqrt{P^\mu_{2i}P^\mu_{2(n-i)}Y_{2i,2(n-i)}}\leqslant\max_{k\geqslant 2}\,\, \sum_{n=2}^k\sqrt{(n+1)\alpha_n}+\sum_{n=k+1}^\infty\sum_{i=0}^n\sqrt{P^\mu_{2i}P^\mu_{2(n-i)}}\\
&\leqslant\max_{k\geqslant 2}\sqrt{\dfrac{(k+4)(k-1)}{2}(Q_d^{\mu,\mu}-\sum_{n=k+1}^{+\infty}\sum_{i=0}^nP^\mu_{2i}P^\mu_{2(n-i)}-\sum_{n=0}^{2}\sum_{i=0}^nP^\mu_{i}P^\mu_{n-i}Y^l_{i,n-i})}+\sum_{n=k+1}^\infty\sum_{i=0}^n\sqrt{P^\mu_{2i}P^\mu_{2(n-i)}},
\end{aligned}
\end{equation}
where, Cauchy-Schwarz inequality is also used.
So the optimization problem Eq.\eqref{cons2} becomes finding $k$ that can maximize the right-hand side of Eq.\eqref{cons4}. So far we get the last line of Eq.\eqref{psiee} in the main text. 
\begin{equation}\begin{aligned}
\label{psiee}
x_{00}&\leqslant \max_{k\geqslant 2}\big|\sqrt{P^\mu_{0}P^\mu_{0}Y^u_{0,0}}+\sqrt{P^\mu_{0}P^\mu_{2}Y^u_{0,2}}+\sqrt{P^\mu_{2}P^\mu_{0}Y^u_{2,0}} + \sum_{n=k+1}^{+\infty}\sum_{i=0}^n\sqrt{P^\mu_{2i}P^\mu_{2(n-i)}}+\\
&\sqrt{\dfrac{(k+4)(k-1)}{2}(Q_d^{\mu,\mu}-\sum_{n=k+1}^{+\infty}\sum_{i=0}^nP^\mu_{2i}P^\mu_{2(n-i)}-\sum_{n=0}^{2}\sum_{i=0}^nP^\mu_{i}P^\mu_{n-i}Y^l_{i,n-i})}\big|^2
\end{aligned}\end{equation}
This method is also effective for estimating upper bounds of $x_{10}$, $x_{01}$ and $x_{11}$. We just post the results below.
\begin{equation}
\begin{aligned}
x_{10}&\leqslant \max_{k\geqslant 1}\big|\sqrt{P^\mu_{1}P^\mu_{0}Y^u_{1,0}}+\sum_{n=k+1}^{+\infty}\sum_{i=0}^n\sqrt{P^\mu_{2i+1}P^\mu_{2(n-i)}}+\\
&\sqrt{\dfrac{(k+3)k}{2}(Q_d^{\mu,\mu}-\sum_{n=k+1}^{+\infty}\sum_{i=0}^nP^\mu_{2i+1}P^\mu_{2(n-i)}-\sum_{n=0}^{2}\sum_{i=0}^nP^\mu_{i}P^\mu_{n-i}Y^l_{i,n-i
})} \big|^2\\
\end{aligned}
\end{equation}

\begin{equation}
\begin{aligned}
x_{01}&\leqslant \max_{k\geqslant 1}\big|\sqrt{P^\mu_{0}P^\mu_{1}Y^u_{0,1}}+\sum_{n=k+1}^{+\infty}\sum_{i=0}^n\sqrt{P^\mu_{2i}P^\mu_{2(n-i)+1}}+\\
&\sqrt{\dfrac{(k+3)k}{2}(Q_d^{\mu,\mu}-\sum_{n=k+1}^{+\infty}\sum_{i=0}^nP^\mu_{2i}P^\mu_{2(n-i)+1}-\sum_{n=0}^{2}\sum_{i=0}^nP^\mu_{i}P^\mu_{n-i}Y^l_{i,n-i
})} \big|^2\\
\end{aligned}
\end{equation}

\begin{equation}
\begin{aligned}
x_{11}&\leqslant \max_{k\geqslant 1}\big|\sqrt{P^\mu_{1}P^\mu_{1}Y^u_{1,1}}+\sum_{n=k+1}^{+\infty}\sum_{i=0}^n\sqrt{P^\mu_{2i+1}P^\mu_{2(n-i)+1}}+\\
&\sqrt{\dfrac{(k+3)k}{2}(Q_d^{\mu,\mu}-\sum_{n=k+1}^{+\infty}\sum_{i=0}^nP^\mu_{2i+1}P^\mu_{2(n-i)+1}-\sum_{n=0}^{2}\sum_{i=0}^nP^\mu_{i}P^\mu_{n-i}Y^l_{i,n-i
})} \big|^2\\
\end{aligned}
\end{equation}

\end{document}